\documentstyle[aps,preprint]{revtex} 
\def\eeq{\end{equation}} 
\def\beq{\begin{equation}} 
\def\bea{\begin{eqnarray}} 
\def\eea{\end{eqnarray}} 
\begin{document} 

\title{On the correct entropic form for systems with 
power-law behaviour: the case of dissipative maps} 

\author{Ramandeep S. Johal$^1$ and Ugur Tirnakli$^{2}$} 

\address{$^1$ Institut f\"{u}r Theoretische Physik,
Technische Universit\"{a}t Dresden, \\ 01062 Dresden, Germany \\
$^2$Department of Physics, Faculty of Science, Ege University, 35100 \\
Izmir, Turkey \\ 
rjohal@theory.phy.tu-dresden.de , tirnakli@sci.ege.edu.tr}  

\maketitle
\begin{abstract}
Maximum entropy principle does not seem to distinguish between 
the use of Tsallis and Renyi entropies as either of them 
may be used to derive similar power-law distributions. 
In this paper, we address the question whether the Renyi entropy 
is equally suitable to describe those systems with power-law behaviour, 
where the use of the Tsallis entropy is relevant. We discuss 
a specific class of dynamical systems, namely, one dimensional 
dissipative maps at chaos threshold and make our study from two aspects: 
i)~power-law sensitivity to the initial conditions and the rate of entropy 
increase, ii)~generalized bit cumulants. We present evidence that the 
Tsallis entropy is the more appropriate entropic form for such studies as 
opposed to Renyi form. 

\noindent {\it PACS Number(s): 05.45.-a, 05.20.-y, 05.70.Ce} 
\end{abstract} 
\newpage 
\section{Introduction} 
As it is evident, an enormous variety of systems in Nature fall into 
the domain of validity of Boltzmann-Gibbs (BG) thermostatistics. On the 
other hand, it is also well-known since long that a variety of anomalous 
systems exists for which the powerful BG formalism exhibits serious 
difficulties. Some of the examples for such anomalous cases could be 
the long-ranged interacting systems \cite{long}, two-dimensional 
turbulence \cite{turbulence}, nonmarkovian processes \cite{nonmarkov}, 
granular matter \cite{granular}, cosmology \cite{cosmo}, high energy 
collisions \cite {highenergy}, among others. In order to handle some of 
these anomalous systems, an attempt has been performed in 
1988 \cite{tsallis88}, which is based on the generalization of the 
standard BG formalism by postulating a nonextensive entropy (Tsallis 
entropy) of the form 
\beq S_q \equiv \frac{1-\sum_{i=1}^W p_i^q}{q-1}\;\; (q\in {\cal R})\;\;, 
\label{st}\eeq and it recovers the standard BG entropy $S_1=-\sum_{i=1}^W p_i \ln p_i$ 
in the $q\rightarrow 1$ limit. This generalization is usually called in 
the literature as Nonextensive thermostatistics or Tsallis 
thermostatistics (TT) \cite{tt}, and it has been (still continues to be!) 
a matter of intensive studies during the past decade \cite{review} both 
from the point of theoretical foundations of the formalism and its 
applications to various physical systems. The apparent success of TT gave 
rise to an increase of the studies with new entropy 
definitions \cite{entropies}. At this point, we would like to mention a 
trivial but important issue: TT by no means intends to cover all of the 
anomalous physical systems where BG statistics seems to fail. It appears 
to be useful only for a large class of cases where power-law behaviour is 
observed. In this sense, alternate forms of entropy could of course 
be useful for other subexponential cases and therefore other possible 
generalizations of the standard formalism could always be projected using 
different entropic forms. One of these attempts, based on the Renyi entropy, 
has become popular nowadays \cite{lenzi,naudts}. The form of this entropy 
is \cite{renyi} 
\beq 
S^R_q \equiv \frac{\ln \sum_i p_i^q}{1-q}. 
\label{sr}\eeq 
It is an extensive quantity (unlike Tsallis entropy), concave for
$0<q<1$ and it recovers the 
usual BG entropy as a special case when $q\rightarrow 1$. 
It is worth mentioning that the Renyi entropy has already been used in 
the multifractal theory \cite{fractal}. On the other hand, the efforts of 
establishing a thermostatistics from this entropy seem to be originated 
from the fact that the maximum entropy principle yields the same form of 
distribution function (of power-law type) for both Tsallis and Renyi 
entropies since they are monotonic functions of each other. At this level, 
there is no a priori reason to choose one of these entropies as the 
correct entropic form for the systems which exhibit power-law type 
behaviour. This is not the case for the standard BG entropy, namely, 
everybody knows that any monotonic function of the BG entropy also gives 
the same distribution function of exponential type, but still the correct 
description of entropy is of course BG. This unclear situation of the 
generalized formalism seems to force some authors, for example Arimitsu 
and Arimitsu, to use unclear statements like "An analytical expression 
of probability density function of velocity fluctuation is derived with 
the help of the statistics based on the Tsallis entropy or the Renyi 
entropy" \cite{arimitsu}. 

From the above discussion, a straightforward question arises naturally: 
Which of these two entropic forms is the correct definition for 
systems exhibiting power-law behaviour ? Our motivation in this paper 
is to try to provide an answer (at least to make the first step) to this 
important and intriguing question. As already seen, since maximum entropy 
principle cannot provide an answer, in order to accomplish this task, 
we shall focus on two different issues and make use of the results coming 
from them. These issues are i)~power-law sensitivity to the initial 
conditions and entropy increase rates, ii)~generalized bit cumulant theory.

\section{Power-law sensitivity to the initial conditions and entropy 
increase rates} 
As is well-known from the theory of dynamical systems (see for example 
\cite{hilborn,beck}), for one-dimensional dissipative systems, it is 
possible to introduce a sensitivity function of type 
\beq 
\xi (t) \equiv \lim_{\Delta x(0)\rightarrow 0} \frac{\Delta x(t)} 
{\Delta x(0)}\;\;, 
\eeq 
where $\Delta x(0)$ and $\Delta x(t)$ are discrepancies of the initial 
conditions at times $0$ and $t$ respectively, and it satisfies the 
differential equation $d\xi/dt=\lambda_1 \xi$, where $\lambda_1$ is the 
standard Lyapunov exponent, thus $\xi(t)$ is of exponential type 
($\xi(t)=e^{\lambda_1t}$). Consequently, when $\lambda_1>0$, the system 
is strongly sensitive to the initial conditions, whereas it is strongly 
insensitive if $\lambda_1<0$. On the other hand, there is an infinite 
number of points for which $\lambda_1=0$. This case is called the marginal 
case and no further analysis of this case is possible within the standard 
theory. It has been conjectured recently that \cite{TPZ} whenever $\lambda_1$ 
vanishes, the sensitivity function becomes of power-law type 
\beq
\xi(t)=\left[1+(1-q)\lambda_q t\right]^{1/(1-q)},
\label{sicq}
\eeq
which is the solution of the differential equation $d\xi/dt=\lambda_q \xi^q$. 
This new definition of the sensitivity function recovers the standard 
exponential one in the $q\rightarrow 1$ limit and here, $\lambda_q$ is the 
generalized Lyapunov exponent and it scales with time inversely as 
$\lambda_1$ does, but this time exhibits a power-law behaviour. 
Consequently, if $\lambda_q>0$ and $q<1$ ($\lambda_q<0$ and $q>1$), then the 
system is weakly sensitive (insensitive) to the initial conditions. 
Most important case which is in the domain of this scenario is of course the 
chaos threshold. At this point (where $\lambda_1=0$ but $\lambda_q\neq 0$), 
the sensitivity function presents strong fluctuations with time and delimits 
the power-law growth of the upper bounds of a complex time dependence of 
the sensitivity to the initial conditions. These upper bounds allow us, 
on a log-log plot, to measure the slope ($\xi(t)\propto t^{1/(1-q)}$) from 
where the proper $q=q^*$ value of the dynamical system under consideration 
can be estimated. This constitutes method I for determining $q^*$ value of a 
dynamical system at the edge of chaos (or at any point where $\lambda_1=0$) 
and it has already been successfully used for a variety of one-dimensional 
dissipative maps such as the standard logistic \cite{TPZ}, 
$z$-logistic \cite{costa}, circle \cite{lyra}, $z$-circular \cite{ugurcircle}, 
single-site \cite{singlesite} and asymmetric logistic \cite{ugurasym} maps. 
At this stage, there is no a priori reason for relating this $q$ index with 
that of the Tsallis entropy. It might well be that the corresponding entropy 
is the Renyi entropy since it also produces the power-law behaviour.

The above concluding statement is also valid for method II of obtaining 
the $q^*$ value of a dynamical system. This method is based on the 
multifractal geometrical aspects of chaotic attractor at the edge of chaos. 
This geometry is characterized by the multifractal singularity spectrum 
$f(\alpha)$, which reflects the fractal dimension of the subset with 
singularity strenght $\alpha$ \cite{fractal,beck}. 
The $f(\alpha)$ function is a down-ward parabola-like concave curve and 
at the end points of this curve, the singularity strenght is associated 
with the most concentrated ($\alpha_{min}$) and the most rarefied 
($\alpha_{max}$) regions of the attractor. The scaling behaviour of these 
regions has been used to propose a new scaling relation as 
\beq 
\frac{1}{1-q^*} = \frac{1}{\alpha_{min}} - 
\frac{1}{\alpha_{max}}\;\; , \label{mulf}
\eeq 
which constitutes a completely different way of calculating the proper 
$q^*$ value of a given dynamical system \cite{lyra}. It is evident from 
the previous works that, for all dissipative systems studied so far, 
the results of these two methods for the proper $q^*$ value coincide 
within a good precision.

Finally, the connection between these $q^*$ values 
and the index $q$ of the Tsallis entropy became evident after the introduction 
of method III for finding $q^*$ values . This method basically makes use of 
a specific generalization of Kolmogorov-Sinai (KS) entropy $K_1$. For a 
chaotic dynamical system, the rate of loss of information can be 
characterized by this entropy and it is defined as the increase of the 
BG entropy per unit time. Since the Pesin equality \cite{pesin} states 
that $K_1=\lambda_1$ if $\lambda_1 >0$ and $K_1=0$ otherwise, it is clear 
that the KS entropy is deeply related to the Lyapunov exponents. Moreover, 
the KS entropy could be defined through 
\beq
K_1\equiv \lim_{t\rightarrow\infty} \lim_{W\rightarrow\infty} 
\lim_{N\rightarrow\infty} \frac{S_1(t)}{t},
\label{k1}
\eeq
where $t$ is the number of discrete time steps in case of maps, $W$ is 
the number of regions in the partition of the phase space and $N$ is the 
number of initial conditions that are evolving with time. Analogously, 
for the marginal cases, a generalized version of the KS entropy $K_q$ has 
been introduced recently \cite{TPZ} by replacing the BG entropy with the 
Tsallis entropy, namely,
\beq
K_q\equiv \lim_{t\rightarrow\infty}
\lim_{W\rightarrow\infty} \lim_{N\rightarrow\infty} \frac{S_q(t)}{t}. 
\label{kq}
\eeq
Consistently, the Pesin equality is expected to become 
$K_q=\lambda_q$ if $\lambda_q >0$ and $K_q=0$ otherwise. 
Using these ideas, the method III \cite{latora,ugurKS} is based on the 
conjecture that (i) a unique value of $q^*$ exists such that $K_q$ is finite 
for $q=q^*$, vanishes for $q>q^*$ and diverges for $q<q^*$; 
(ii) this $q^*$ value coincides with that coming from methods I and II. 
Latora et al. have examined numerically the standard logistic map 
\beq 
x_{t+1} = 1 - a\; x_t^2 \;\; , 
\eeq 
(where $0<a\le 2$ and $-1\le x_t \le 1$) both in the chaotic region and at 
the chaos threshold \cite{latora}. In the chaotic region (for example, 
$a=2$ case), they numerically verified that $K_q$ vanishes (diverges) for 
any value of $q>1$ ($q<1$), being finite only for $q=1$ 
(see Fig.1 of \cite{latora}), which implies that the proper $q^*$ value is 
unity as expected. On the other hand, at the chaos threshold, the same 
structure has also been observed, but this time with an important 
exception: $K_q$ vanishes (diverges) for any value of $q>q^*$ ($q<q^*$), 
being finite only for $q=q^*=0.24...$ (see Fig.4 of \cite{latora}). 
This value coincides with the value of the proper $q^*$ coming from other 
two methods. This structure has been verified not only for the standard 
logistic map but also a variety of other map families as 
well \cite{singlesite,ugurasym,ugurKS}.

Now we are ready to proceed with our main purpose in this effort. 
We conjecture here that if the Tsallis and Renyi entropies are both the 
correct choices for a system with power-law behaviour, then the Renyi 
entropy should also give us the proper $q^*$ values whenever we use this 
entropic form instead of the Tsallis entropy in method III. 
Before demonstrating our results, let us describe the numerical procedure: 
Firstly, we partition the phase space into $W$ equal cells, then we choose 
one of these cells and select $N$ initial conditions all inside the chosen 
cell. As time evolves, these $N$ points spread within the phase space and 
it gives a set $\{N_i(t)\}$ with $\sum_{i=1}^W N_i(t)=N$, $\forall t$, 
which consequently yields a set of probabilities from where one can 
calculate the entropy. In order to compare our results to those of 
Latora et al., we numerically check the Renyi entropy case for the 
standard logistic map both in the chaotic region and at the chaos threshold. 
The results are illustrated in Fig.1. Surprisingly, the expected 
structure has not been observed. For the chaotic region (Fig.1a), it is 
seen that, for $q=1$ case, the linear entropy increase is realized but 
very strangely other values of $q$ also give the same result! This shows that
Renyi entropy is a much {\it less} sensitive function as compared
to Tsallis entropy with respect to changes in the value of $q$.
Similarly, at the chaos threshold (Fig.1b), Renyi entropy does not allow
us to distinguish the correct $q^*$ value for which a linear rate
is expected. Note that rate of increase of Renyi entropy can be
linear if we plot the entropy with respect to ${\rm ln}(t)$ variable,
which follows due to the relation between Tsallis and Renyi entropy, 
(compare (\ref{st}) and (\ref{sr})).
This is another reason for unsuitability of the Renyi entropy; for Tsallis
entropy we get the result of strong chaos (BG entropy proportional 
to $t$) in the $q\to 1$ limit, but for Renyi entropy, $q\to 1$ 
implies that BG entropy is proportional to ${\rm ln}(t)$. 
These observations strongly 
suggest that the Renyi entropy is not the suitable entropic form for these 
dynamical systems which exhibit power-law behaviour at their marginal points 
(where $\lambda_1=0$). Although the results which led us to make this 
claim seem rather convincing, it would be no doubt convenient to seek 
further evidences which can strengthen this claim. This is what we shall 
try to do in the remainder of the paper by making use of the generalized 
bit cumulants.

\section{Generalized bit cumulants} 
Bit statistics is a tool to describe the complicated
probability distributions, such as generated by
chaotic systems \cite{beck}. In this framework, the first cumulant
is the BG entropy itself, the second cumulant
gives the variance (a measure of fluctuations) 
of the 'classical' bit number
$b_i = -{\rm ln}\;p_i$, and so on.
Particularly, the second bit cumulant which is a generalization 
of specific heat, can be applied in
the context of equilibrium and non-equilibrium phase transitions
\cite{shlogl}. The classical bit cumulants can be generalized 
\cite{rrrj} within TT formalism. 
They were applied to symmetric logistic and $z$-logistic 
family of maps given by 
\beq
x_{t+1} = 1-a|x_t|^z,
\label{symap}
\eeq
where the inflexion parameter $z>1$, $0<a\le 2$ and
$-1\le x\le 1$ and $t =0,1,2,...$. Applications have also 
been found for asymmetric family of logistic maps 
\cite{utasym}, given by

\begin{eqnarray}
x_{t+1} = \left\{ 
\begin{array}{ll}
1- a|x_t|^{z_1}, & \mbox{if $x_t \ge 0$} \\ 
1- a|x_t|^{z_2}, & \mbox{if $x_t \le 0$}
\end{array} \right\},
\label{asymap}
\end{eqnarray}
where $z_i>1$. We briefly summarise the motivation and also 
the information obtained from these applications, as we intend to 
apply this approach to elucidate the difference in the use of 
Renyi and Tsallis entropies.

The generalized second bit cumulant, which gives variance of the
Tsallis bit number $-[a_i]_q \equiv {(p_i)^{q-1}-1\over {1-q}}$, is given by
\beq
C^{(T)}_{2} = {1\over (1-q)^2}\left[\sum_{i} (p_i)^{2q-1}-
(\sum_{i} (p_i)^{q})^2\right].
\label{ct2}
\eeq
For $q\to 1$, (\ref{ct2}) approaches the standard bit variance, 
\beq
C_2 = \sum_{i}({\rm ln}\;p_i)^2 p_i
- (\sum_i p_i{\rm ln}\;p_i)^2.
\label{stc2}
\eeq
It was observed in \cite{rrrj} that the ratio 
$C^{(T)}_{2}/C_2$ evaluated in the chaotic region 
of the map (\ref{symap}), gives a scaling factor referred to as 'slope' 
in this paper (see Figs.), whose value depends on values of $q$ as well as the 
inflexion parameter $z$. Naturally, the slope tends to unity for $q\to 1$. 
Less trivially, it was observed that the slope tends to unity 
also for $z\to \infty$. In other words, increasing $z$ value within 
the map-family, has the same effect on the 
quantifier 'slope' as the effect of taking to unity, the $q$ parameter 
entering in its definition. Thus the otherwise free parameter $q$ in (\ref{ct2})
behaves analogous to the proper $q^*$ value, which shows a monotonic
decrease with increasing $z$ values \cite{costa}. 
This observation was used to conjecture the 
behaviour of proper index $q^*$ versus $(z_1,z_2)$ pairs \cite{utasym} 
in maps (\ref{asymap}) at chaos threshold. The actual behaviour 
obtained from finding $q^*$ values for these maps by using the 
previously introduced methods \cite{ugurasym}, agree with the 
conjecture of \cite{utasym}. This implies that 
generalized second bit cumulant (\ref{ct2}) can be 
consistently applied in such an analysis. 

Can we have generalization of bit-cumulants based on the Renyi entropy ?
If such a possibility exists, then these new quantities
may be used to study the implications on the above mentioned
maps, and a comparison be made of the predictions derived from
Tsallis based bit-cumulants and Renyi based ones.
This task of a new generalization is not a straightforward one,
in view of the fact that there is no generalized bit number corresponding
to the Renyi entropy, i.e., the latter cannot be written as usual
mean over some bit numbers, unlike the cases of BG and Tsallis entropies.
In fact, Renyi entropy can be expressed only as a kind of nonlinear 
average \cite{kn}. Recently, however, a step was made in this direction by
one of the authors \cite{rjdd}. Specifically, the discrete
derivative operator can be used to generate two types of
bit cumulants, say type {\rm A} and type {\rm B}. 
Type {\rm A} cumulants are Tsallis type, of the kind formulated in \cite{rrrj}. 
These are nonextensive quantities with respect to independent 
subsystems and preserve the standard relations between bit-moments and
cumulants, as do the cumulants and moments of a probability 
distribution for a certain random variable. Type {\rm B} 'cumulants'
are however extensive quantites but do not preserve the 
standard relations between bit-moments and cumulants.
Still, the approach allows us to clearly distinguish the separate 
origins of Tsallis and Renyi entropies,
based on the use of discrete derivative. Particularly, the first
bit cumulant of type {\rm B} is the Renyi entropy, the second bit cumulant 
is another positive valued $q$-generalization of the standard bit variance,
\beq
{C}^{(R)}_{2} = {1\over (1-q)^2}\left[{\rm ln}\;\sum_{i} (p_i)^{2q-1}
- 2{\rm ln}\;\sum_{i} (p_i)^{q} \right],
\label{cr2}
\eeq
It also recovers the form (\ref{stc2}) in the limit 
$q\to 1$. Thus it is interesting to see the application
of (\ref{cr2}) analogous to that of (\ref{ct2}) for the 
1-{\rm d} dissipative maps. We expect that any difference
in the predictions based on (\ref{cr2}) as compared to
those of (\ref{ct2}), also reflects the separate
implications for Renyi and Tsallis entropies for such systems. 

In the following, we compare the results 
obtained from using the generalized second bit-cumulants 
(\ref{ct2}) and (\ref{cr2}). In Fig. 2a, the results for 
symmetric maps (\ref{symap}) are given. The slope which implies 
the ratio $C^{(R)}_{2}/C_2$, shows a non-monotonic behaviour 
with increasing $z$ values. This can be contrasted with 
the monotonic decrease of the ratio $C^{(T)}_{2}/C_2$, as 
shown in Fig. 2b. Similarly, for asymmetric maps (\ref{asymap}), 
we see separate trends as is clear by comparing Figs. 3a and 3b. 

This makes us to conclude that while the $q$ parameter 
in the definition of generalized cumulant $C^{(T)}_{2}$ 
behaves like the $q^*$ parameter inferred from the 
sensitivity to initial conditions at chaos threshold (method I), 
the $q$ parameter in the definition $C^{(R)}_{2}$ 
serves no such connection. Thus this generalization 
of the standard cumulants, which are related to Renyi entropy, 
are inappropriate for such studies. 
This clearly strengthens our claim that the Renyi entropy is not 
the appropriate entropic form for such dynamical systems exhibiting 
power-law behaviour.

\section{Rate of entropy increase revisited}
So far, we have argued and shown that 
although it is expected to be applicable to systems 
exhibiting power law behaviour, Renyi entropy is not a 
suitable measure to quantify the rate of entropy increase at points of 
power-law sensitivity in 1-d maps. However, 
Tsallis entropy qualifies as the appropriate measure 
for this purpose. A natural question follows: Is Tsallis 
entropy the unique measure in this regard ? 
In this section, we propose that Tsallis entropy is {\it not} the 
unique measure of entropy which can give a linear rate 
of increase at chaos threshold. We exemplify below by 
proposing an alternative nonextensive entropy which 
can also yield a linear rate of increase. However, our conjecture is that 
the Tsallis entropy seems to be the unique entropy which yields the same $q^*$
as obtained from the modified sensitivity function (\ref{sicq}) and 
multifractal spectrum (\ref{mulf}), i.e., methods I and II to 
determine $q^*$. To understand this role of the Tsallis entropy, 
we have to appreciate the implication of Pesin's equality.

First, let us discuss the case of strong exponential sensitivity 
$\lambda_1 >0$. In (\ref{k1}) assuming equiprobability, 
$p_i =1/M(t)$, where $M(t)$ are the number of microstates 
or cells occupied at time $t$, we have 
\beq
K_1 = \lim_{t\rightarrow\infty} \lim_{W\rightarrow\infty}
\lim_{N\rightarrow\infty} {1\over t} {\rm ln}\;M(t).
\label{km}
\eeq
Standard Pesin's equality $K_1=\lambda_1$ allows us to relate 
the above expression with exponential sensitivity function. 
As a result, we get 
\beq
\xi(t) = M(t).
\label{delm}
\eeq
In other words, for strong exponential sensitivity 
the number of states increase linearly with the distance between 
trajectories for all times $t>0$. 

When $\lambda_1 =0$, the use of modified sensitivity function 
(\ref{sicq}) is suggested. Then we have $\lambda_q >0$. 
Using the definition (\ref{kq}) of $K_q$ and 
assuming equiprobability, we get 
\beq
K_q =\lim_{t\rightarrow\infty}
\lim_{W\rightarrow\infty} \lim_{N\rightarrow\infty}
{1\over t} {M(t)^{1-q} -1\over 1-q}. 
\label{kqm}
\eeq
Now postulating the generalized Pesin equality for the proper $q^*$
value of the entropic index (which yields the linear rate
of entropy increase), $K_{q^*}=\lambda_{q^*}$, 
we get from (\ref{sicq}) and (\ref{kqm}), the same relation as (\ref{delm}) 
even for points of power-law sensitivity. 
Thus the use of the Tsallis entropy in the definition of 
$K_q$ ensures that the linear relation (\ref{delm}) 
is preserved. Finally, this automatically implies 
that the critical $q^*$ value obtained in methods I and III are identical. 

Now consider an alternate entropic form 
\beq
{\cal S}_q = \frac{1-\sum_{i} p_i^{(1+{\rm ln}\;q)}}{q-1}\;\; 
(q\in {\cal R}^+)\;\;,
\label{cals}
\eeq
which has the same nonextensive property as the Tsallis entropy. 
It approximates the Tsallis entropy for values of $q$ 
very near to one, and thus goes to BG entropy 
in the limit $q\to 1$. 
For this entropy, we can also obtain a linear rate 
of increase at chaos threshold, for certain critical value 
of $q=q_c$. But in this case, $q_c\ne q^*$, where $q^*$ is 
calculated from methods I and II. Instead, we have 
$1+{\rm ln}\;q_c = q^*$. A more thorough understanding 
of generalized Pesin's equality is required to relate
such other entropies with the issue of power-law sensitivity.
At the moment, we only state that there are other 
entropic forms which can give a linear rate of increase, but 
the entropic form compatible with the generalized sensitivty
function (\ref{sicq}), in the sense that methods I and III
should yield same $q^*$ values, is the Tsallis entropy.

\section{Conclusions} 
In this study, the appropriate entropic form for systems exhibiting 
power-law behaviour has been tried to be determined. To accomplish 
this task, two different issues have been analysed. 
One of these issues is the sensitivity to initial conditions and 
the entropy increase rates, the other one is the generalized bit 
cumulants. From the investigation of both issues, for the dynamical 
systems at marginal points (where the standard Lyapunov exponent 
vanishes), it is verified that the suitable entropic form is the 
Tsallis entropy. This could be considered as the first step towards 
the answer of the intriguing question: What is the correct entropic 
form for systems exhibiting power-law behaviour ?

The reasons for the success of the Tsallis entropy, we believe, 
are two-fold: i)~Tsallis entropy as function of time $t$, 
reproduces the correct 
$q\to 1$ limit of BG entropy proportional to $t$, 
ii)~Tsallis entropy is a much more sensitive function than
Renyi entropy with respect to changes in $q$ value, which
helps to determine the proper $q^*$ value in method III.
A crucial property of Tsallis entropy is
its nonextensivity. It is not certain whether this
property plays significant role in such studies. As has been
argued in \cite{wada}, a conditional entropy which 
is nonextensive for generic values of $q$, becomes 
extensive for the proper value of $q=q^*$. It seems that, for such 
kind of systems with power-law behaviour, what is important is NOT 
to be extensive for all $q$ values, but rather, to be extensive for 
only a special value of $q$ (namely $q^*$), the value that gives the 
correct entropy for the system under investigation. 

Finally, it is worth mentioning that we also show that the Tsallis 
entropy is not the unique entropic form which gives a linear entropy 
increase rate, but on the other hand, it seems to be the unique one which 
provides $q^*$ values consistent with those obtained from the other 
two methods (i.e., from the sensitivity function and from the 
multifractal scaling relation).

\section*{Acknowledgments} 
RSJ is grateful to Alexander von Humboldt Foundation, Germany,
for financial support.

\newpage 
{\bf Figure Captions} \\ 

{\bf Figure 1} - Time evolution of the Renyi entropy at the chaotic 
region (a) and at the chaos threshold (b) for various $q$ values. 
Inset of (b): The behaviour of the Renyi entropy for ${\rm ln}t$ .\\ 

{\bf Figure 2} - The behaviour of the slope as a function of the 
map inflexion parameter $z$ for the Renyi (a) and Tsallis (b) cases. \\

{\bf Figure 3} - The behaviour of the slope as a function of the 
map inflexion parameter pairs $(z_1,z_2)$ for the Renyi (a) and 
Tsallis (b) cases. \\ 

\newpage 

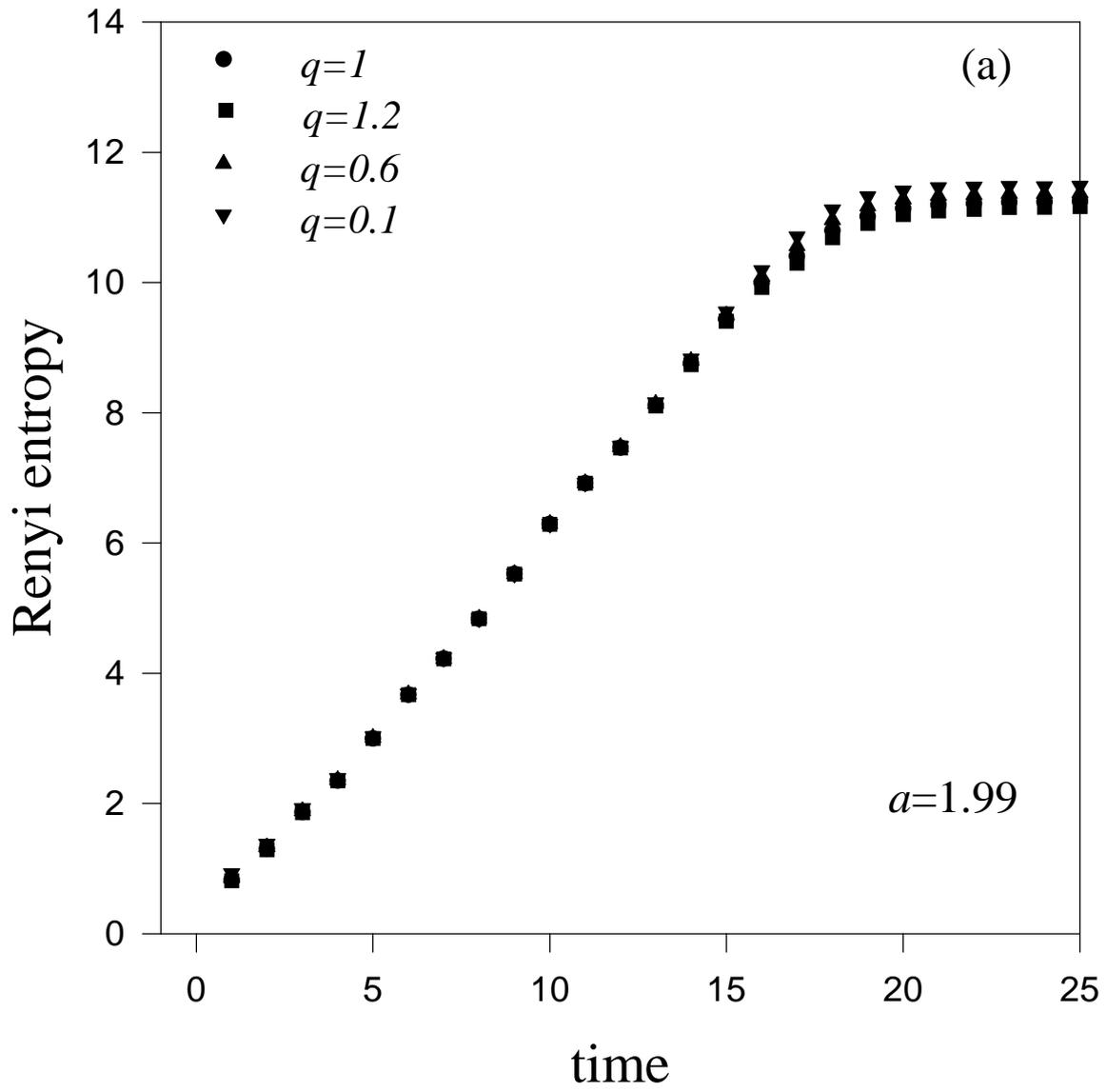
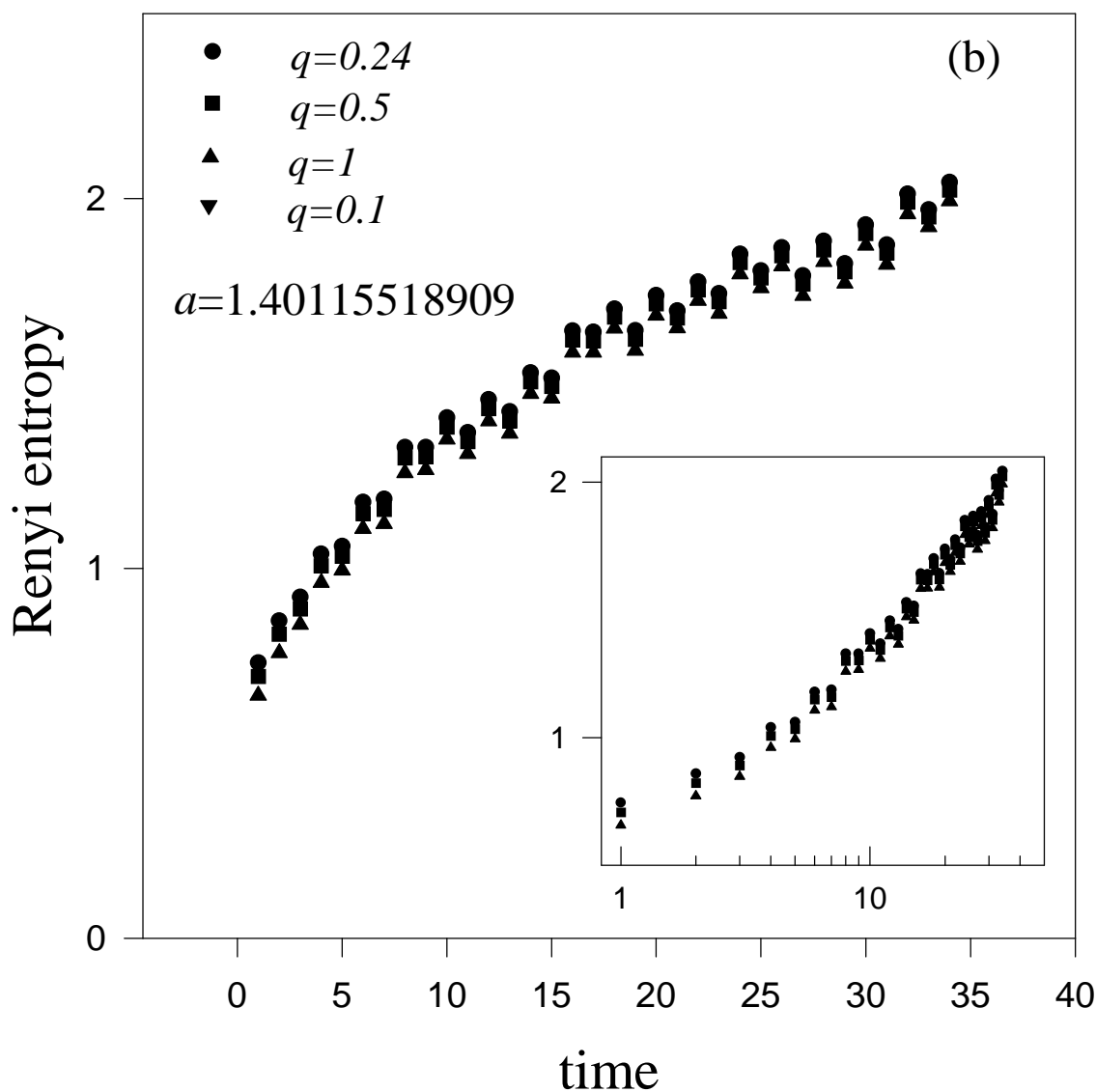
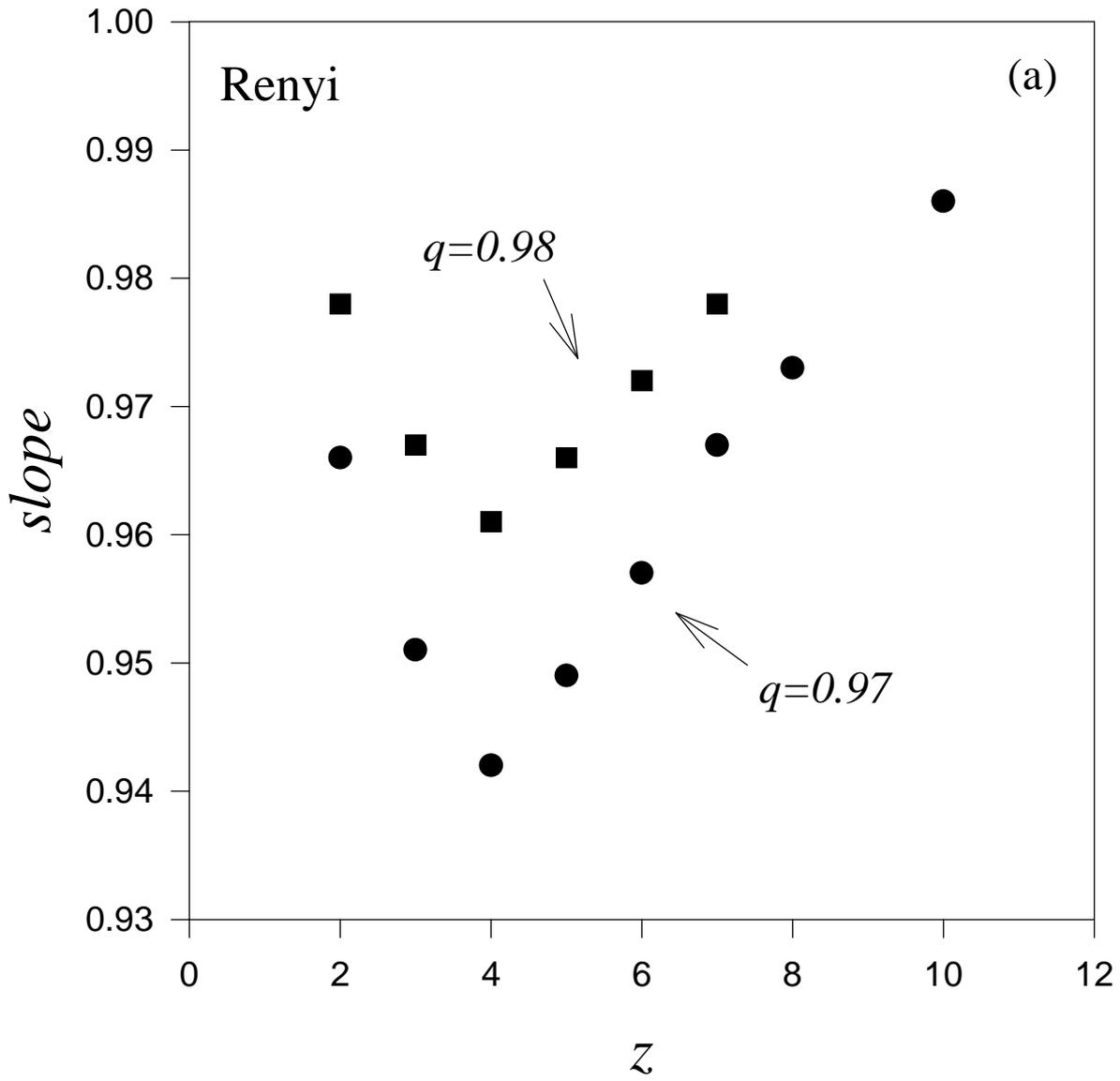
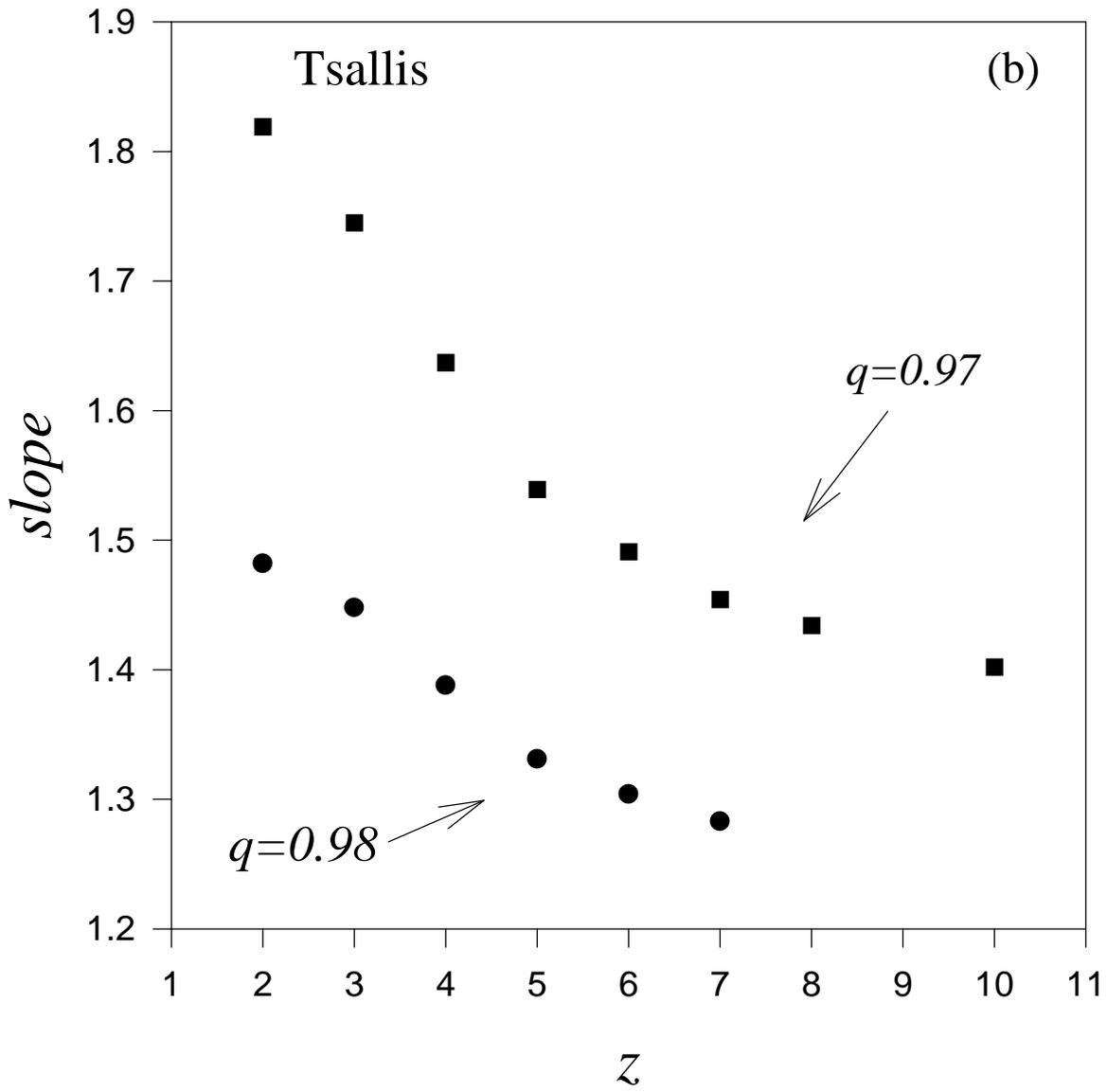
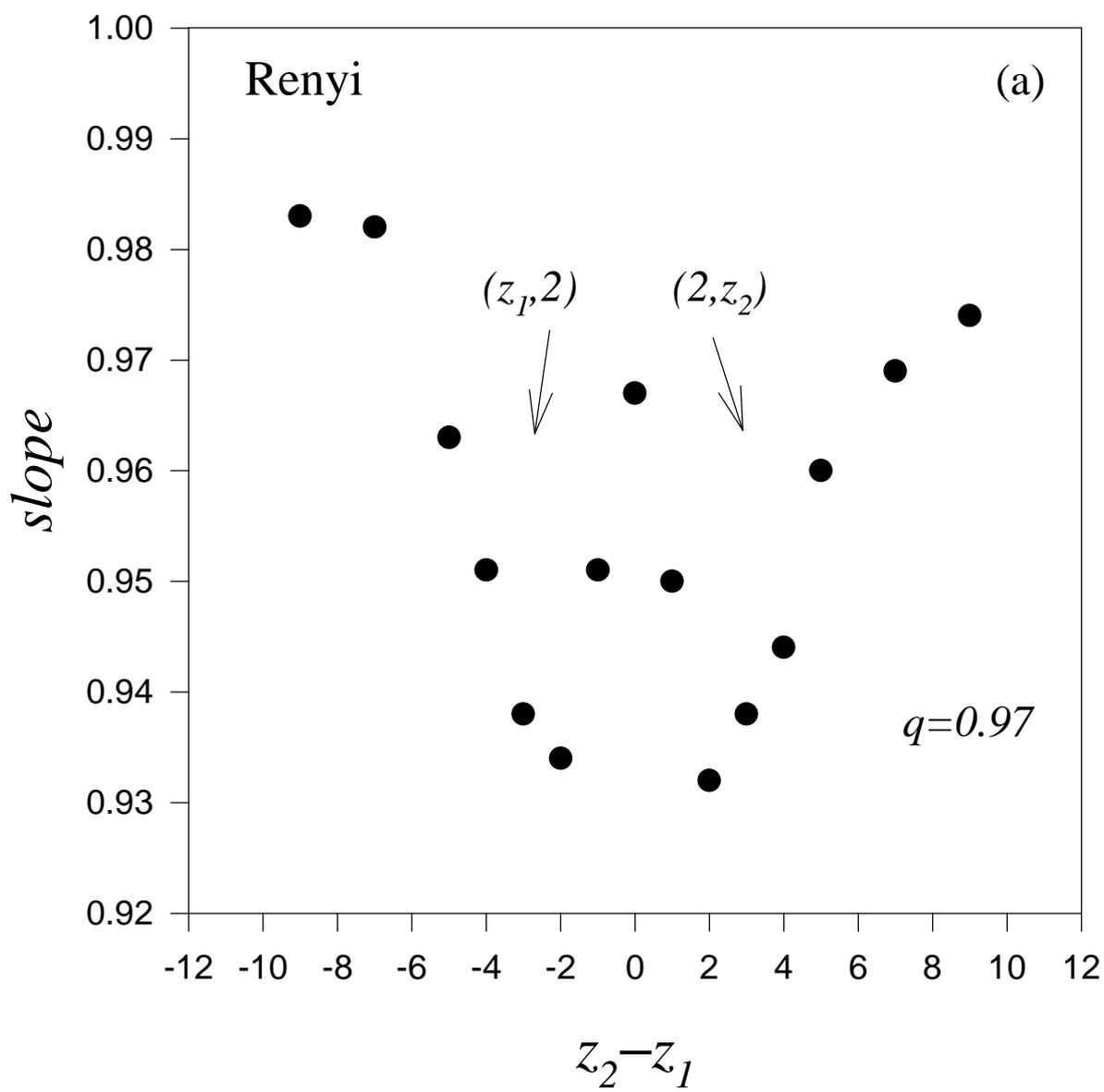
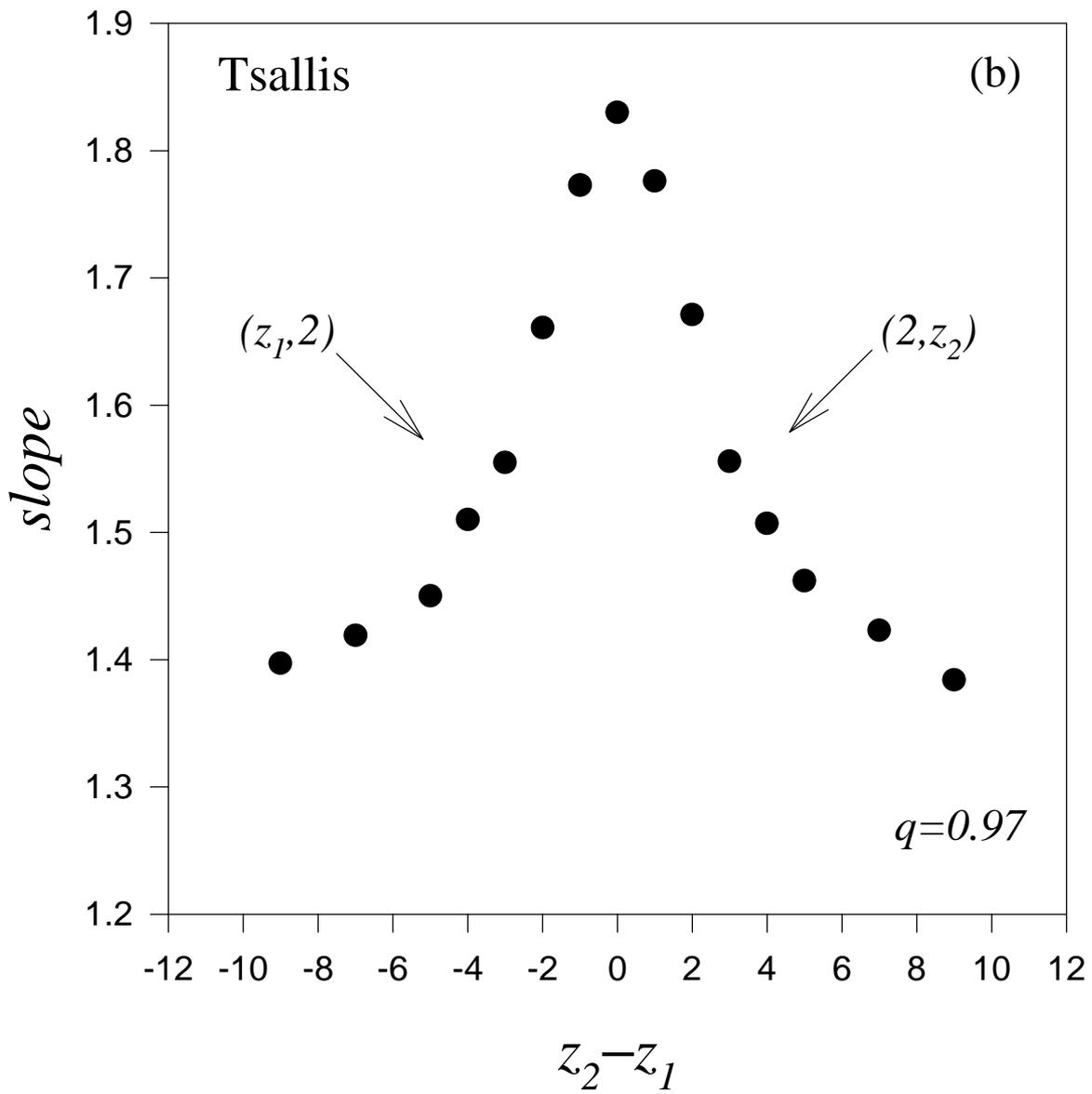
\end{document}